\documentclass[12pt,a4paper]{article}
\usepackage[T1]{fontenc}
\usepackage[utf8]{inputenc}
\usepackage{graphicx}
\usepackage{amsmath,amssymb}
\usepackage{mathrsfs}
\usepackage{authblk}
\usepackage[noadjust]{cite}
\usepackage{hyperref}
\hypersetup{
    unicode=false,
    pdftitle={Universal spectral correlations in ensembles of random normal matrices}, 
    pdfauthor={Ravi Prakash, Akhilesh Pandey},
    pdfnewwindow=true,
    colorlinks=true, 
    linkcolor=red,
    citecolor=blue,
    filecolor=magenta,
    urlcolor=cyan
}
\title{\textbf{Universal spectral correlations in ensembles of random normal matrices}}
\author{Ravi Prakash \thanks{Email: raviprakash.iit@gmail.com}}
\author{Akhilesh Pandey \thanks{Email: ap0700@mail.jnu.ac.in}}
\affil{School of Physical Sciences, Jawaharlal Nehru University, New Delhi-110067, INDIA}
\date{}
\begin{document}
\maketitle
\begin{abstract}
We consider non-gaussian ensembles of random normal matrices with the constraint that the ensembles are invariant under unitary transformations. We show that the level density of eigenvalues  exhibits disk to ring transition in the complex plane. We also show that the $n$-eigenvalue correlation and the spacing distribution are universal and identical to that of complex (Gaussian) Ginibre ensemble. Our results are confirmed by Monte Carlo calculations. We verify the universality for dissipative quantum kicked rotor system.
\end{abstract}

\section{Introduction}
Random matrix ensembles play prominent role in the study of quantum systems which exhibit chaotic behavior in the classical limits. The local spectral and transition width fluctuations as well as scattering cross section and conductance fluctuations have been extensively studied and are well described by random matrix ensembles \cite{rmp-apandey, bgs-2, rmp-beenakker, guhr,stockmann, mlmehta, oxford}. Ensembles of hermitian  matrices (e.g. Gaussian orthogonal, unitary or symplectic ensemble) and of unitary matrices  (e.g. circular orthogonal, unitary or symplectic ensemble) are widely used.

Ginibre introduced ensemble of general complex matrices with no unitary or hermitian condition imposed \cite{ginibre}. The matrix elements follow the Gaussian distribution, i.e., the joint probability distribution (jpd) for Ginibre matrices $G$ is given by $P(G) \propto \exp[-Tr~V(G^\dag G)]$, where $V(x) = x$ with $x > 0$. Eigenvalues lie in the complex plane and are distributed uniformly in a disc centered at origin \cite{ginibre, mlmehta}. The Ginibre ensemble and its extension to non-gaussian weight functions (viz. nonlinear potentials $V$) have found applications in several fields e.g., quantum dissipative systems with eigenvalues having imaginary parts due to dissipation \cite{qdissipation,qdissipation2,haakebook, d_braun}, quantum chromodynamics at finite chemical potential \cite{ markum, markum2}, transition from symmetric to asymmetric matrices \cite{sommers,fyodorov}, neural network dynamics \cite{neural}, truncation of scattering matrices \cite{mesoscopic} and google networks \cite{google}.

Ginibre ensembles with non-gaussian weight are typically difficult to dealt with because of lack of symmetry.  We consider ensembles of normal matrices with arbitrary weight functions. Normal matrices commute with their hermitian adjoints. Such ensembles are exactly solvable because the matrices are diagonalizable by unitary transformations \cite{g-oas}. We believe that normal matrix ensembles will capture the behavior of the spectral density qualitatively and the behavior of two-eigenvalue and higher order correlations exactly. Moreover, normal matrix ensembles are useful in the study of fractional quantum Hall effect where the Laughlin wave function for non-uniform electromagnetic \cite{fqhe-1, fqhe-2} field is similar to the eigenvalue jpd and also in study of other problems, where the matrix may not be normal but eigenvalue jpd is similar \cite{mesoscopic, fishmann}.

Our purpose in this Letter is to study ensembles of normal matrices with arbitrary weight. We show that the eigenvalues are scattered in the complex plane such that it can constitute either a disk or a ring structure for arbitrary potential \cite{single_ring}. We study the two-eigenvalue and higher order correlations as well as spacing distribution to show that unfolded spectra are stationary and universal. We confirm our results by Monte Carlo (MC) simulation \cite{gpps} of several non-gaussian ensembles. As an application, we also study effect of dissipation on the quantum kicked rotor maps \cite{izrailev, pramana-kicked-rotor, pragya-cue}. In the non-dissipative chaotic regime, the evolution operator is equivalent to circular random matrix ensemble \cite{pramana-kicked-rotor, pragya-cue}. The dissipation term is modeled by a non-unitary random matrix perturbation. We show that the eigenvalue distribution for both, the kicked rotor and the random matrix model are qualitatively similar. We also obtain the same universal correlations for both cases as in the case of normal matrix models.

\section{Joint probability distribution and correlation functions}
We consider complex normal matrices $M$ with jpd of matrix elements given by,
\begin{equation}
\label{matrix-jpd}
\mathcal{P}(M) \propto e^{- N~Tr\left[V\left(M^\dag M\right)\right]}.
\end{equation}
The eigenvalues of a normal matrix can be obtained by a unitary transformation $U$, i.e., $M$ can be written as
\begin{equation}
\label{matrix-transformation}
M  = U Z U^\dag,
\end{equation}
where $Z \equiv \mbox{diag}\left(z_1,\ldots,z_N\right)$. The $z_j$ are the eigenvalues of matrix $M$ and are in general complex. The jpd of eigenvalues can be obtained by transformation from matrix space to eigenvalue space followed by integral over the eigenvector variables. Thus
\begin{equation}
\label{jpd-transformation}
P(z_1,\ldots,z_N) = \int \ldots \int \mathcal{P}\left(U Z U^\dag\right) J(z,U) \mu(\mathrm d U),
\end{equation}
where $\mu$ denotes the infinitesimal volume element of $U$ and $J$ denotes the Jacobian of the transformation from $M$ to $(z,U)$ variables. Integrating over $U$ we get \cite{g-oas,mlmehta},
\begin{equation}
\label{ev-jpd}
P(z_1,\ldots,z_N) = C \prod\limits_{j<k}|z_j - z_k|^\beta e^{-N \sum\limits_{l=1}^N V\left(|z_l|^2\right)},
\end{equation}
where $C$ is the normalization constant and $\beta = 2$. We mention that $\beta = 1$ and $4$ correspond to the complex normal matrices which are symmetric and self-dual respectively. In these cases the matrices are diagonalizable by orthogonal and symplectic transformations respectively. In this paper, we consider $\beta = 2$ case. We remark that the jpd in (\ref{ev-jpd}) for $\beta = 2$ turns out to be identical to the jpd for non-normal ensembles for the Gaussian weight \cite{mlmehta, ginibre} and some logarithmic \cite{fishmann, mesoscopic} potentials. The jpd is rotationally invariant in the complex plane. The factor $N$ in the exponent will ensure that for large $N$, any disk or ring structure has $N$-independent radii. We will be interested in the $n$-eigenvalue correlation functions, $R_n(z_1,\ldots,z_n)$, defined by
\begin{equation}
R_n = \frac{N!}{(N-n)!}\int \ldots \int P(z_1,\ldots,z_N)~\mathrm d^2 z_{n+1}\ldots\mathrm d^2 z_N,
\end{equation}
where $\mathrm d^2 z$ represents the infinitesimal area element, i.e., $\mathrm d^2 z = \mathrm d x \mathrm dy$ in Cartesian coordinates. Here $R_1$ is the spectral density. To derive the $n$-eigenvalue correlation functions we follow the method of orthogonal polynomials in the complex plane. Following \cite{ginibre, mlmehta} in the Gaussian case and \cite{g-oas} for general potential, we introduce the orthogonal polynomials, $p_l(z) = z^l$, where,
\begin{equation}
\label{orthogonality}
\int z^l \left(z^\star\right)^{l^\prime} e^{-N V\left(|z|^2\right)} ~\mathrm d^2 z  = N_l \delta_{ll^\prime},
\end{equation}
with $N_l$ the normalization constant. Note that the orthogonality in (\ref{orthogonality}) comes from the angular integration. We define the kernels,
\begin{equation}
\label{two-point-kernel}
K_N\left(z_j,z_k\right) = e^{-\frac{N}{2}\left[V\left(|z_j|^2\right) + V\left(|z_k|^2\right)\right]} \sum\limits_{l=0}^{N-1} \frac{\left( z_j z_k^\star \right)^l}{N_l}.
\end{equation}
Then n-eigenvalue correlation functions are given by \cite{ginibre, mlmehta, g-oas}
\begin{equation}
\label{n-level-correlation}
R_n\left(z_1,\ldots,z_n\right) = \det\left[K_N\left(z_j,z_k\right)\right]_{j,k=1,\ldots,n}.
\end{equation}
Thus for example
\begin{equation}
\label{r1-kernel}
R_1(z) = K_N(z,z),
\end{equation}
and, since $K_N\left( z_j, z_k \right) = K_N^\star \left(z_k,z_j \right)$, we have,
\begin{equation}
\label{r2-kernel}
R_2\left(z_1,z_2\right) = R_1\left(z_1\right) R_1\left(z_2\right) - \left| K_N\left(z_1,z_2\right)\right|^2.
\end{equation}
\section{Large $N$ expression for spectral density}
\label{sec-spectral-density}
The jpd of eigenvalues is equivalent to the partition function of $N$ particles in the complex plane under the interaction of repulsive two-body log potential and a one-body binding potential $V$. We have
\begin{equation}
\label{partition-function}
P\left(z_1,\ldots,z_N\right) = C e^{-W},
\end{equation}
where the effective potential $W$ is given by
\begin{equation}
\label{effective-potential}
W = -\sum_{1\leq j < k \leq N} \ln\left|z_j - z_k\right|^2 + N \sum_l V\left(|z_l|^2\right).
\end{equation}
The jpd will be maximum when $W$ is minimum, i.e., $\partial W/\partial z_j = 0$ and $\partial W/\partial z_j^\star = 0$. Thus for large $N$,
\begin{equation}
\label{equation-4}
 \sum_{k(\neq j)} \frac{1}{z_j - z_k} = z_j^\star N V^\prime \left(\left|z_j\right|^2\right).
\end{equation}
Here $V^\prime$ is the differential with respect to its argument $|z|^2$. For large $N$ the summation in (\ref{equation-4}) can be replaced by an integral over $R_1$. Using isotropic invariance of $R_1$, we find from (\ref{equation-4}) that,  with $R_1(z) = R_1(|z|)$,
\begin{equation}
 \int \frac{R_1(|\xi|)}{z - \xi} ~\mathrm d^2 \xi = z^\star N V^\prime\left(\left|z^2\right|\right).
\end{equation}
Defining $\xi = r \exp(i \theta)$ and integrating over $\theta$, we get,
\begin{equation}
\label{first_case}
\int_0^{|z|}r R_1(r)~\mathrm dr = \frac{|z|^2}{2\pi}N V^\prime \left(|z|^2\right).
\end{equation}
Differentiating with respect to $|z|$, we have finally
\begin{equation}
\label{r1}
 R_1(|z|) = \frac{N}{\pi} \left[V^\prime \left(|z|^2\right) + |z|^2 V^{\prime \prime}\left(|z|^2\right)\right].
\end{equation}
Note that the left hand side of (\ref{first_case}) is greater than zero and monotonically increasing function of $|z|$. Therefore the right hand side must show the same behavior. Thus all the eigenvalues must lie in the range where $V^\prime(|z|^2)$ is positive and monotonically increasing. This implies that whatever the potential, the constraint allows eigenvalues to lie either in a disk shape or in a ring shape only. This is referred to as the single ring theorem \cite{single_ring}. The inner radius, $a$, is obtained from,
\begin{equation}
\label{r_in}
V^\prime\left(|z|^2\right) \geq 0~\mbox{and monotonic for} |z| \geq a.
\end{equation}
The outer radius, $b$, is obtained from the normalization condition,
\begin{equation}
\label{r_out}
 2\pi\int_a^b r R_1(r) ~\mathrm dr = N.
\end{equation}
If there are several solutions of $(\ref{r_in})$, we choose the largest value of $a$. See the example of cosine potential ahead.

\section{Spectral density for potentials with arbitrary boundaries}
\label{sec-spectral-density-bounded}
We have assumed above that the potential is smooth and its first and second derivatives are continuous. We now consider hard walls at $|z_1|$ and $|z_2|$, where $|z_1| < |z_2|$. Then there are several scenarios. If $|z_1| \leq a < b \leq |z_2|$, the result (\ref{r1}) remains the same. If $a \leq |z_1| < |z_2| \leq b$, then the density in (\ref{r1}) is modified. In this case the eigenvalues up to $|z_1|$ accumulate at $|z_1|$ and eigenvalues beyond $|z_2|$ accumulate at $|z_2|$. This behavior is consistent with the condition (\ref{first_case}) and is verified for the Gaussian potential ahead. In this case the spectral density is defined for $|z_1| \leq |z| \leq |z_2|$ and is given by,
\begin{align}
\label{r1-bounded}
\nonumber R_1(|z|) &= \frac{N}{\pi} \left[ |z_1| V^\prime\left(\left|z_1\right|^2\right) \delta\left(|z| - \left|z_1\right|\right)     \right. \\
\nonumber & + |z|^2 V^{\prime \prime}\left(|z|^2\right) + V^\prime\left(|z|^2\right) \\
&\left. + \left(\frac{1}{\left|z_2\right|} - \left|z_2\right|V^\prime\left(\left|z_2\right|^2\right)\right)\delta(|z| - \left|z_2\right|)\right].
\end{align}
Other cases can be similarly dealt with.

\section{Universality of two-eigenvalue and higher order correlations}
To evaluate $n$-eigenvalue correlation functions we require the knowledge of the normalization constant in (\ref{two-point-kernel}) for arbitrary potential. From (\ref{two-point-kernel},\ref{r1-kernel}) along with (\ref{r1}), we find for large $N$,
\begin{equation}
\label{r1-oas-largeN}
e^{-N V\left(|z|^2\right)} \sum\limits_{l=0}^{N-1} \frac{|z|^{2l}}{N_l} = \frac{N}{\pi} \left[V^\prime\left(|z|^2\right) + |z|^2 V^{\prime \prime}\left(|z|^2\right)\right].
\end{equation}
Thus the normalization constant $N_l$ is given by,
\begin{equation}
\label{nl}
\frac{1}{N_l} =  \frac{N}{\pi l!} \frac{d^l}{dx^l} \left[ \exp\left(N V(x)\right)   \left[V^\prime(x) + x V^{\prime \prime}(x)\right] \right]_{x \rightarrow 0}.
\end{equation}
Substituting (\ref{nl}) in (\ref{two-point-kernel}) we get for large $N$
\begin{align}
\label{kernel}
 K_N\left(z_j,z_k\right)=& \frac{N}{\pi} e^{-\frac{N}{2} \left(V\left(|z_j|^2\right) +  V\left(|z_k|^2\right)\right)}
   \left[z_j z_k^\star V^\prime(z_j z_k^\star)\right]^\prime e^{N V(z_j z_k^\star)} .
\end{align}
Eq.(\ref{kernel}) gives analytic expression for the two point kernel for arbitrary potential $V$.

In order to investigate the universality of correlation functions we have to transform the spectrum so that the transformed spectral density is uniform. This procedure is called unfolding of the spectrum. For the Gaussian Ginibre ensemble the density is uniform  and therefore unfolding is not required. We choose the new density to be $1/\pi$ so that we can compare with the results of \cite{mlmehta, ginibre}. Transformation from $z$ variables to new variables $\zeta$ should be such that,
\begin{equation}
\label{equation-5_2}
\frac{1}{\pi} ~\mathrm d^2 \zeta = R_1(|z|) ~\mathrm d^2 z.
\end{equation}
Instead of unfolding the entire spectrum by one function, we consider local unfolding. We assume that the spectral density is constant locally for large $N$. The transformation of line element can be written as,
\begin{equation}
\label{equation-5_3}
\mathrm d \zeta = \sqrt{\pi R_1(|z|)} ~\mathrm d z.
\end{equation}
Thus  $\zeta_j$ is related to $z_j$ in the neighborhood of $z$ by the relation,
\begin{equation}
\label{unfolding}
z_j = z + \delta_j;~\delta_j = \frac{\zeta_j}{\sqrt{\pi R_1(|z|)}},
\end{equation}
where $\zeta_j = O(1)$ and $\delta_j = O(1/\sqrt N)$.

The $n$-eigenvalue correlation function, $\mathcal{R}_n(\zeta_1,\ldots, \zeta_n)$, in transformed variables can be obtained from,
\begin{equation}
\label{equation-7}
\mathcal{R}_n~\mathrm d^2 \zeta_1\ldots \mathrm d^2 \zeta_n = R_n\left(z_1,\ldots,z_n\right) ~\mathrm d^2 z_1 \ldots \mathrm d^2 z_n.
\end{equation}
It follows that $\mathcal R_n$ can be written as,
\begin{equation}
\label{equation-8}
\mathcal R_n\left(\zeta_1,\ldots,\zeta_n\right) = \det\left[\mathcal K_N\left(\zeta_j,\zeta_k\right)\right]_{j,k=1,\ldots,n},
\end{equation}
where
\begin{equation}
\label{mathcal-K_N}
\mathcal K_N \left(\zeta_j,\zeta_k\right) = \frac{1}{\pi}\frac{K_N\left(z_j,z_k\right)}{\sqrt{K_N\left(z_j,z_j\right) K_N\left(z_k,z_k\right)}}.
\end{equation}
Note that $\mathcal R_1 = 1/\pi$, as it should be. Using (\ref{kernel}) in (\ref{mathcal-K_N}) and keeping leading terms we get,
\begin{equation}
\label{kernel-nongauss-1}
\mathcal K_N \left(\zeta_j,\zeta_k\right) = \frac{1}{\pi} \exp\left[-\frac{N}{2} \left(V\left(|z_j|^2\right) + V\left(|z_k|^2\right) - 2V\left(z_j z_k^\star\right)\right)\right].
\end{equation}
Substituting (\ref{unfolding}) in (\ref{kernel-nongauss-1}) and evaluating the exponent to order $N\delta$ and $N\delta^2$, we get,
\begin{align}
\label{kernel-nongauss-2}
\nonumber \mathcal K_N \left(\zeta_j,\zeta_k\right) &= \frac{1}{\pi}\exp\left[ - \frac{\pi}{2} R_1(|z|) \left( |\delta_j - \delta_k|^2 \right)\right] \\
&\times  \exp \left[ \frac{\pi}{2} R_1(|z|) (\delta_j \delta_k^\star - \delta_j^\star \delta_k) + \left( \chi_j- \chi_k \right)\right].
\end{align}
Here $\chi$ is purely imaginary and is given by
\begin{equation}
\chi_j = \frac{N}{2}V^\prime z^\star \delta_j  + \frac{N}{4}V^{\prime \prime} {z^\star}^2 \delta_j^2 - \mbox{C.C.},
\end{equation}
with C.C. denoting the complex conjugate. The $\chi$ term in (\ref{kernel-nongauss-2}) does not affect the determinant in (\ref{equation-8}) and therefore we ignore it. The kernel can then be written in terms of $\zeta$ as,
\begin{equation}
\label{kernel-nongauss}
\mathcal K_N(\zeta_j,\zeta_k) = \frac{1}{\pi} e^{-\frac{1}{2}|\zeta_j - \zeta_k|^2 + \frac{1}{2}(\zeta_j \zeta_k^\star - \zeta_j^\star \zeta_k)}.
\end{equation}
Thus the two-eigenvalue correlation function is given by,
\begin{equation}
\label{r2-nongauss}
\mathcal R_2 (\zeta_1,\zeta_2) = \frac{1}{\pi^2}\left( 1 - e^{-|\zeta_1 - \zeta_2|^2}\right).
\end{equation}
Note that, unlike (\ref{r2-nongauss}), the higher order functions will depend on the phase term of (\ref{kernel-nongauss}).

The result in (\ref{r2-nongauss}) is known for the Gaussian Ginibre case \cite{mlmehta}. Our proof shows that not only (\ref{r2-nongauss}) but the more general $n$-eigenvalue correlation functions given in (\ref{equation-8}, \ref{kernel-nongauss}) are all \textit{stationary} (i.e., independent of $z$) and \textit{universal} (i.e., independent of $V$). Using rotational invariance of the correlation functions, we introduce,
\begin{equation}
\label{r2-nongauss-s}
\mathscr R_2(s) = \iint \mathcal{R}_2(\zeta_1,\zeta_2)\delta(s - |\zeta_1 - \zeta_2|) ~\mathrm{d}^2\zeta_2.
\end{equation}
Substituting eq.(\ref{r2-nongauss}) in (\ref{r2-nongauss-s}), we get for large $N$,
\begin{equation}
\label{r2}
\mathscr{R}_2(s) = \frac{2}{\pi} s \left(1 - e^{-s^2}\right).
\end{equation}
Eq. (\ref{r2}) for small $s$ provides another proof of cubic repulsion \cite{haakebook}. $\mathscr R_2$ can also be written as,
\begin{equation}
\label{r2-as-ps}
\pi \mathscr R_2(s) = \sum_{k=0}^\infty P_k(s),
\end{equation}
where $P_k(s)$ is the density of the $k^\text{th}$ nearest neighbor spacing, $s$. Note that since the $\mathcal R_n$ are stationary and universal and so will be the $P_k(s)$. Thus for example $P_0(s)$ given for the Gaussian ensemble \cite{haakebook},
\begin{equation}
\label{spacing-haake}
P_0(s) = - \frac{d}{ds}\prod_{n=1}^\infty \left[e_n(s^2) e^{-s^2}\right],
\end{equation}
where $e_n(x) = \sum_{j = 0}^n x^j/j!$, will be applicable to all potentials.

\section{Numerical Study}
\label{sec-numerical-results}
We verify the spectral density and two-eigenvalues correlation function numerically using Monte Carlo (MC) simulation for various potentials. In order to obtain samples of spectra from the jpd given in (\ref{ev-jpd}), we extend the MC method developed in \cite{gpps} to complex eigenvalues. We obtain spectral density and correlation functions for several non-gaussian potentials discussed below. For calculation of $\mathscr R_2$, we consider histogram of all spacings suitably normalized as in (\ref{r2-as-ps}). The spacing between two eigenvalues $z_j$ and $z_k$ is unfolded as $s = \left| z_j - z_k\right|\sqrt{\pi R_1((z_j+z_k)/2)}$.

\section{Quartic potential and its correlation functions}
\label{sec-quartic-potential}
\begin{figure}[!h]
\centering
\includegraphics[width=0.7\linewidth]{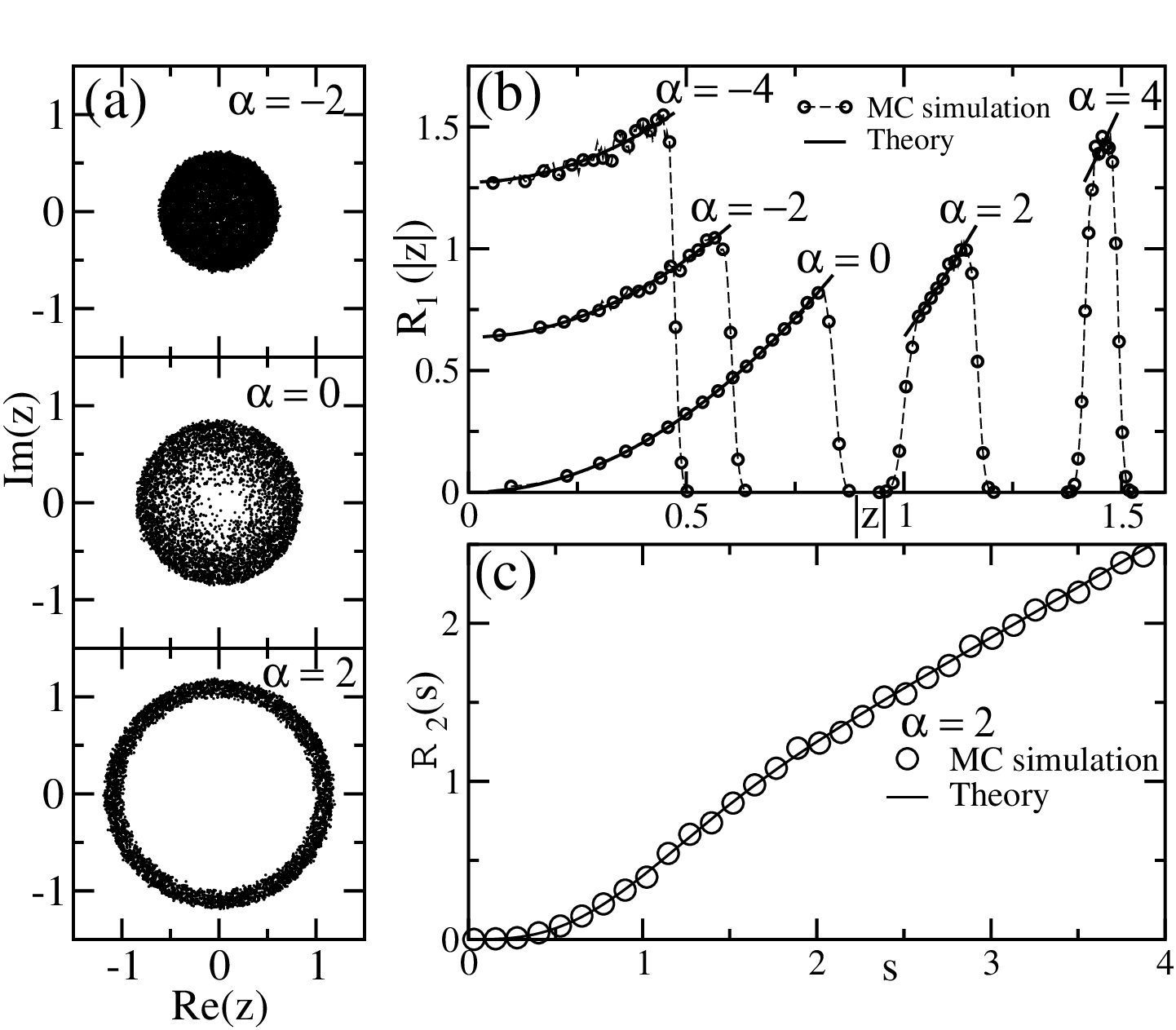}
\caption{\label{ev-r1-r2-2well}(a) Plot of eigenvalues in the complex plane, (b) spectral density, $R_1(|z|)$, (c) two-eigenvalue correlation, $\mathscr R_2(s)$, for the quartic potential with various values of $\alpha$ and $N = 500$.}
\end{figure}

We consider
\begin{equation}
\label{quartic-pot}
V(|z|^2) = |z|^4 - \alpha |z|^2.
\end{equation}
For $\alpha < 0$, this gives a single well potential whereas for $\alpha \geq 0$ it gives a double well potential. Substituting eq.(\ref{quartic-pot}) in (\ref{r1}), we get,
\begin{equation}
\label{density_doublewell}
\nonumber R_1(|z|) = \frac{N}{\pi}\left(4|z|^2-\alpha\right)~\mbox{for}~a \leq |z| \leq b,
\end{equation}
where $a$ and $b$ are inner and outer radii respectively which are obtained by conditions (\ref{r_in}) and (\ref{r_out}). We get $a = (\alpha/2)^{1/2}$ for $\alpha \geq 0$ and $0$ for $\alpha < 0$. Similarly $b = (\alpha/4 + (1/2 + \alpha^2/16)^{1/2})^{1/2}$. Note that there is a disk to ring transition at $\alpha = 0$. Complex eigenvalues obtained by MC simulations and spectral density for various values of $\alpha$ are shown in fig.(\ref{ev-r1-r2-2well}a,\ref{ev-r1-r2-2well}b). Note that the spectral density is not constant.
The two-eigenvalue correlation is shown in fig.(\ref{ev-r1-r2-2well}c) and is consistent with (\ref{r2}).

\section{Logarithmic potential}
\begin{figure}[!h]
\centering
\includegraphics[width=0.8\linewidth, trim=0mm 0mm 0mm 12mm]{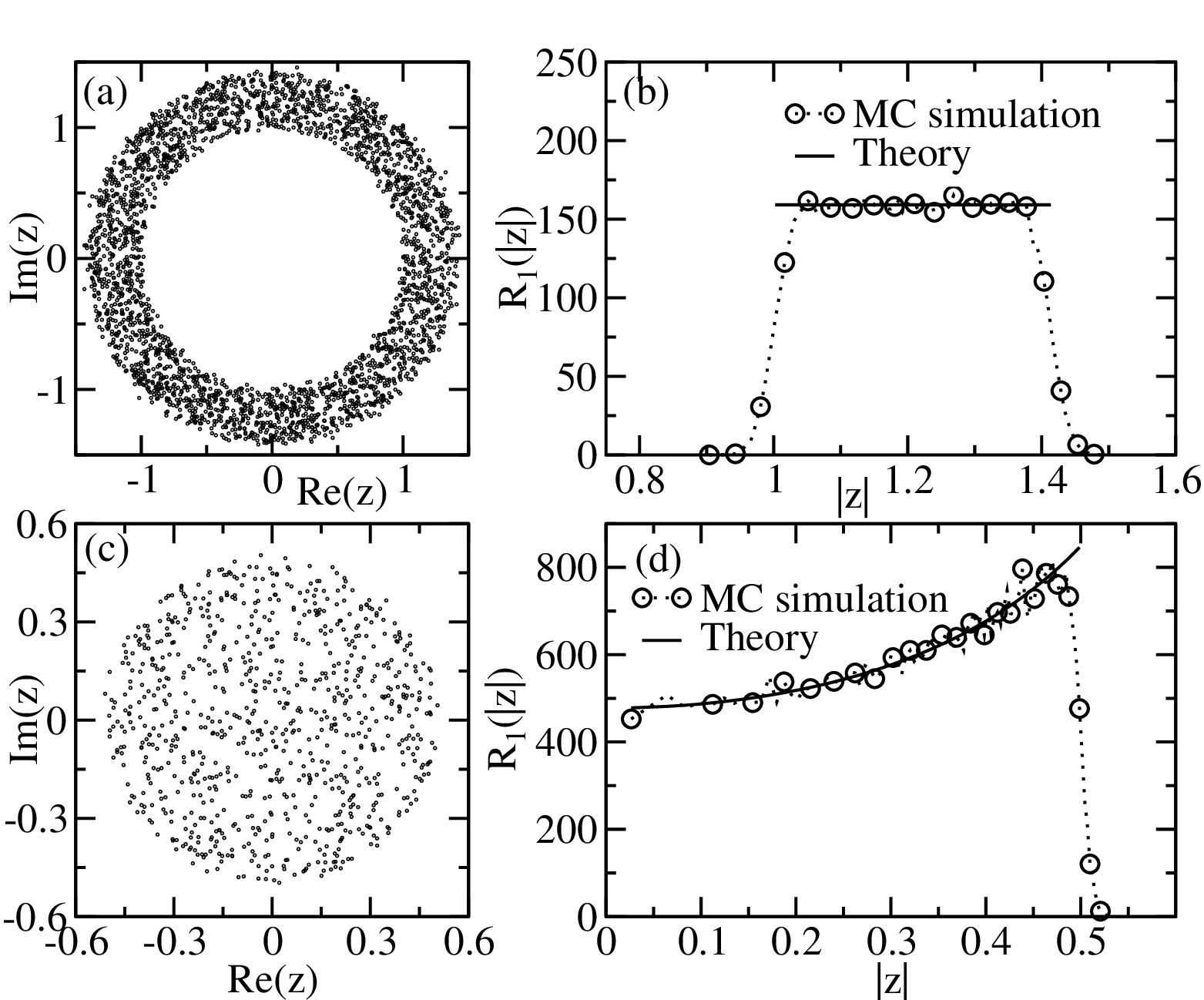}
\caption{\label{ev-log}(a) Eigenvalue plot and (b) spectral density for $V = |z|^2 - \log(|z|^2)$ and similarly plots for truncated log potential, $V = -3 \log(1-|z|^2)$, are shown in (c) and (d) respectively. $N = 500$ in both cases.}
\end{figure}

The Disc to ring transition is also obtained for the logarithmic potential, viz., $V\left(|z|^2\right) = |z|^2 - \alpha \log\left(|z|^2\right)$ for $\alpha \geq 0$ \cite{fishmann}. The eigenvalue density is $R_1(|z|) = N/\pi$ for $a \leq |z| \leq b$ with $a = \sqrt \alpha$ and $b = \sqrt{\alpha + 1}$. See fig.(\ref{ev-log}) for MC verification of $R_1$. $\mathscr R_2(s)$ is similar to the case shown in fig.(\ref{ev-r1-r2-2well}c).

\section{Truncated log potential}
In \cite{mesoscopic} the jpd, spectral density and the two-eigenvalue correlation function for truncated unitary matrices are studied. The jpd is same as (\ref{ev-jpd}) with the potential, $V\left(|z|^2\right) = -\mu \log\left(1-|z|^2\right),~|z| < 1$. The density is $R_1(|z|) = N \mu/\pi(1-|z|^2)^2$ with $a = 0$ and $b = 1/\sqrt{1+\mu}$ (see fig. (\ref{ev-log})) and the $\mathcal R_2$ is as in (\ref{r2-nongauss}).

\section{Cosine potential}
\label{sec-gaussian-bounded-cosine}
\begin{figure}[!h]
\centering
\includegraphics[width=0.75\linewidth, trim=0mm 0mm 0mm 12mm,clip]{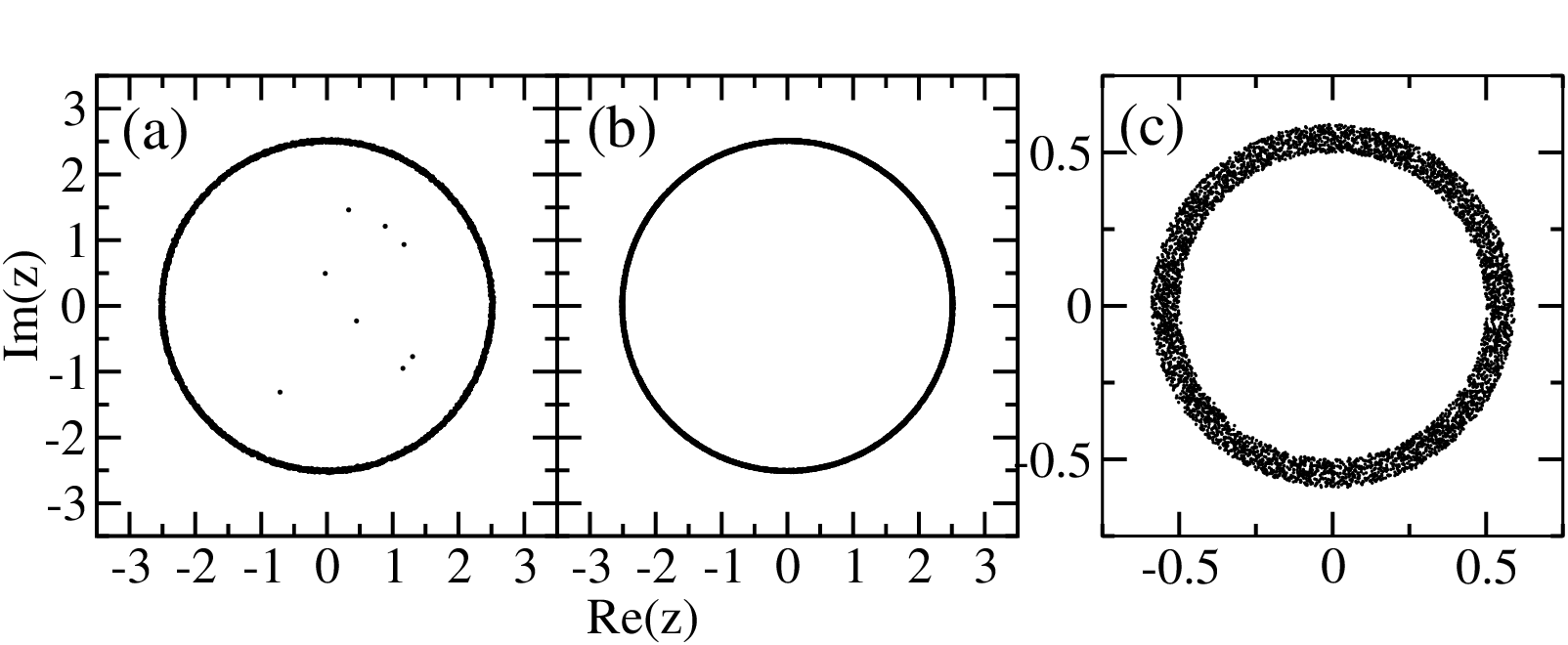}
\caption{\label{ev-3circle}Plot of eigenvalues for $V = \cos(2\pi |z|)- 1$ for (a) $N =500, \gamma = 3$, (b) $N = 20000, \gamma = 3$ and (c) $N = 20000, \gamma = 1$.}
\end{figure}

We consider the cosine potential
\begin{equation}
V = (\cos(n\pi |z|) - 1), ~ |z| < \gamma,
\end{equation}
where $n$ is an integer. The potential is infinite for $|z| \geq \gamma$. The spectral density and radii of the ring can be obtained using eqs.(\ref{r1},\ref{r_in},\ref{r_out}). Eigenvalue plot in the complex plane is shown in fig.(\ref{ev-3circle}). We observe several rings for small values of $N$. As shown in fig.(\ref{ev-3circle}a), there are three rings for $\gamma = 3$ at $N = 500$. However as $N$ increases, eigenvalues move towards the outer most ring. This is seen in fig.(\ref{ev-3circle}b) and supports the single ring prediction for large $N$ case. Same result is shown for $\gamma = 1$ in fig.(\ref{ev-3circle}c).

\section{Gaussian potential with boundaries}
\begin{figure}[!h]
\centering
\includegraphics[width=0.6\linewidth]{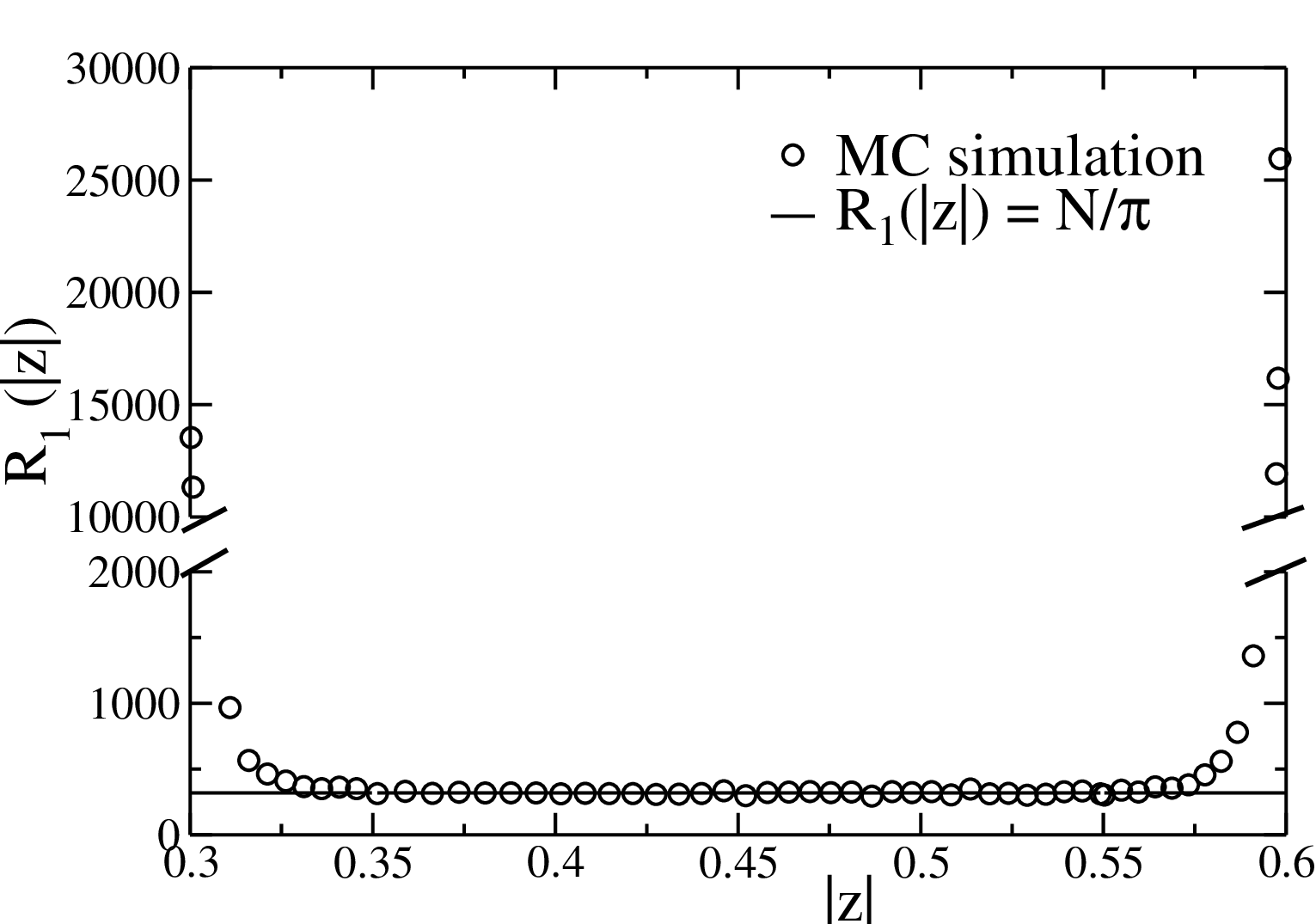}
\caption{\label{fig-ginibre-bounded}Spectral density for Gaussian potential bounded between boundaries place at $|z_1| = 0.3$ and $|z_2| = 0.6$. Spectra of size $N = 1000$ are considered here.}
\end{figure}

We know that eigenvalues are distributed uniformly $(R_1(|z|) = N/\pi)$ in a disc of unit radius for potential $V =  |z|^2$. We consider the same potential with boundaries, $V = |z|^2,  |z_1| < |z| < |z_2|$. The spectral density for this case is shown in fig.(\ref{fig-ginibre-bounded}). The density remains uniform in $\left(|z_1|,|z_2|\right)$ but increases rapidly at the boundaries, as predicted in (\ref{r1-bounded}).

\section{Spacing Distribution}
\label{sec-spacings}
\begin{figure}[!h]
\centering
\includegraphics[height=0.5\linewidth]{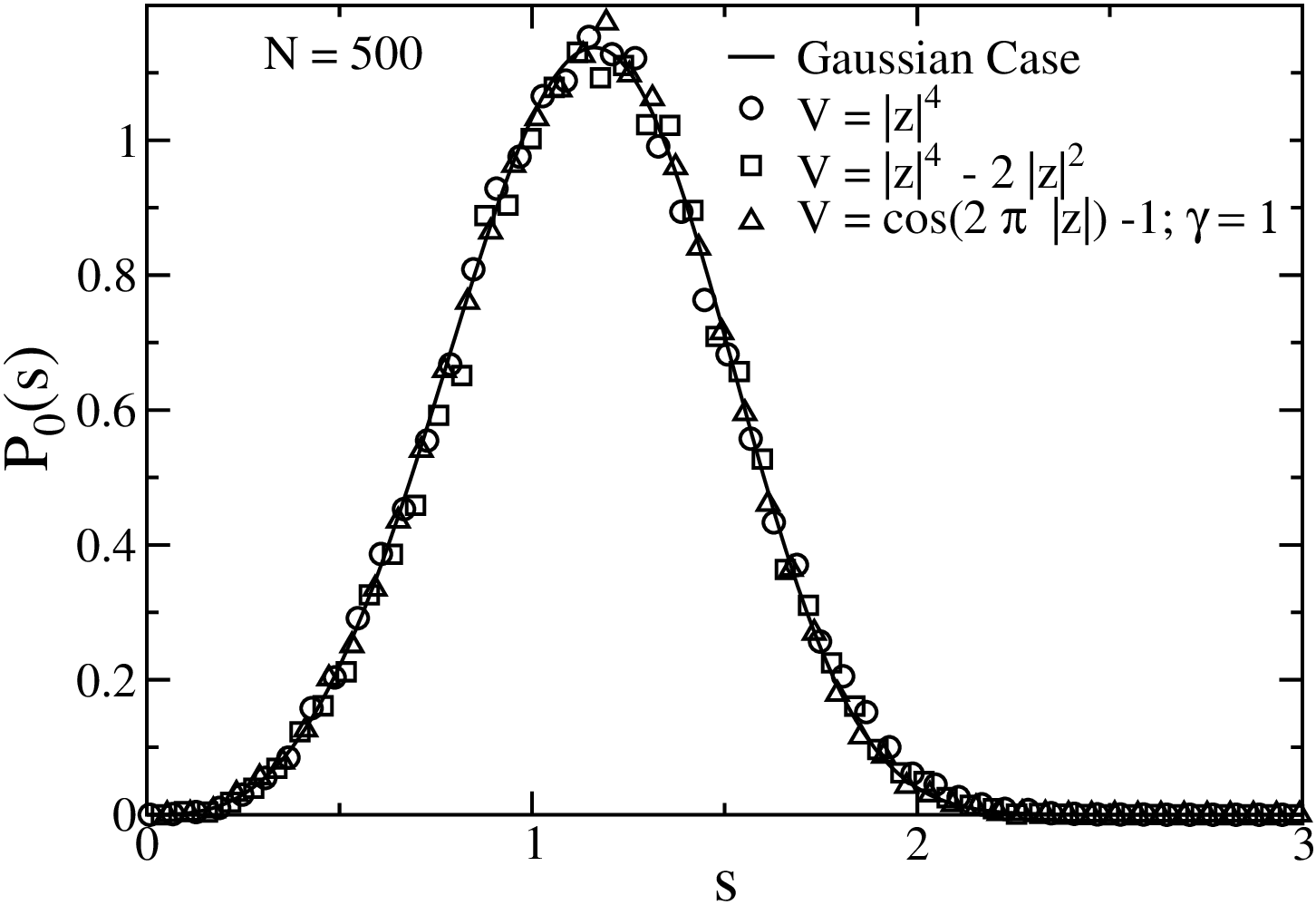}
\caption{\label{sp-nongauss} Nearest neighbor spacing distribution for several non-gaussian potentials.  The smooth line shows the spacing distribution for Gaussian Ginibre ensemble given by (\ref{spacing-haake}).}
\end{figure}

The nearest neighbor spacing distribution $P_0(s)$ is shown in fig.(\ref{sp-nongauss}) for three potentials. The agreement with the theoretical result (\ref{spacing-haake}) is excellent.

\section{Dissipative quantum kicked rotor}
\label{sec-dissipative-kicked-rotor}
\begin{figure}[!h]
\centering
\includegraphics[width=0.7\linewidth]{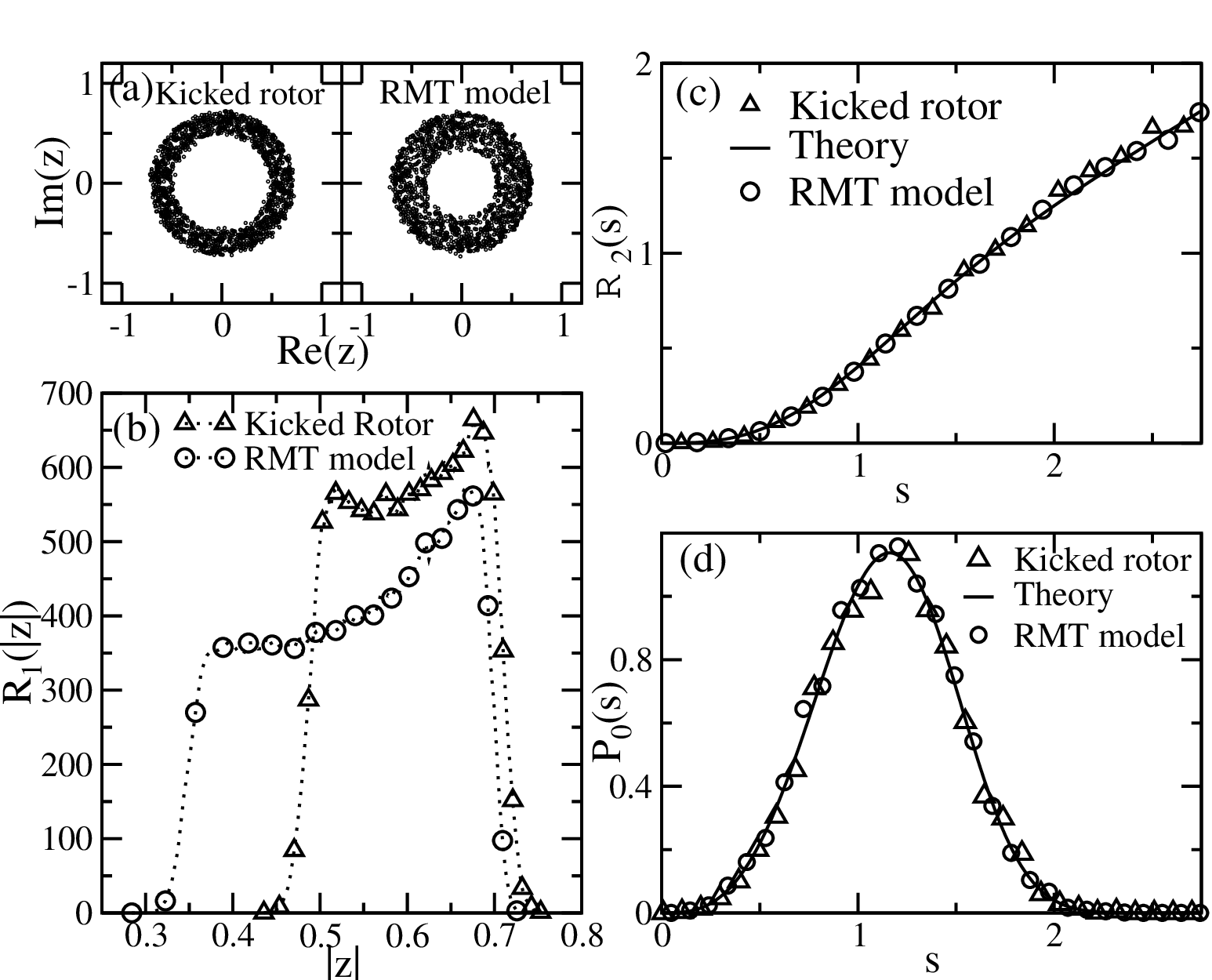}
\caption{\label{ev-r1-r2-sp-krotor-qdmodel} (a) Eigenvalue plot in the complex plane, (b) spectral density, (c) two-eigenvalue correlation and (d) spacing distribution for kicked rotor and the equivalent random matrix theory (RMT) model for $N = 501$.}
\end{figure}

We now consider the effect of dissipation on quantum chaotic maps. We use quantum maps in an $N$-dimensional Hilbert space generated by the time evolution operator \cite{izrailev,pramana-kicked-rotor} of a periodically kicked rotor with torus boundary conditions. For the standard map, the evolution operator is given by $U = B G$, where $B \equiv B(\kappa) = \exp[-i\kappa \cos(\theta + \theta_0)/\hbar]$ and $G = \exp[-i(p + \gamma)^2/2\hbar]$ with $\theta,p$ the position and momentum operators respectively. Here $\kappa$ is the kicking parameter, $\theta_0$ is the parity breaking parameter and $\gamma$ is the time-reversal breaking parameter. For large $\kappa$, the corresponding classical system displays chaos and the quantum system follows the circular ensemble models \cite{izrailev,pramana-kicked-rotor,pragya-cue}. For dissipation, we consider the evolution operator $F(\alpha) = BGD$ where $D \equiv D(\alpha) = \exp[-\alpha p^2/2\hbar]$, and $\alpha (> 0)$ is the dissipation parameter. The matrix elements of $F(0)$ in the position representation are given in \cite{izrailev,pramana-kicked-rotor,pragya-cue}. We obtain similarly the matrix elements of $F(\alpha)$. We have,
\begin{align}
\nonumber & F_{mn}(\alpha) = \frac{1}{N} \exp\left[-i\frac{\kappa}{\hbar}\cos(\frac{2\pi m}{N} + \theta_0)\right] \times \\
& \sum_{l=-N_1}^{N_1} \exp\left[-i\left(\frac{\hbar - i \alpha}{2}l^2 - \gamma l + 2\pi l \left(\frac{m-n}{N}\right)\right)\right]
\end{align}
where $m,n = -N^\prime, -N^\prime + 1,\ldots,N^\prime, N^\prime = (N-1)/2$. We set $\hbar = 1$. If there is no dissipation (i.e. $\alpha = 0$), the evolution operator is unitary (i.e. $F(0) = U$) and eigenvalues lie on circle of unit radius centered at origin. In presence of dissipation, the operator is no longer unitary and eigenvalues start coming inside the circle to constitute a ring like structure. We numerically analyze the eigenvalue spectrum. We take $\theta_0 = \pi/2N$ and $\gamma = 0.7$ to ensure that the parity and time-reversal symmetries are broken and the circular unitary ensemble $(CUE)$ results are obtained for $\alpha = 0$. The eigenvalue plot, two-eigenvalue correlation function and the spacing distribution are shown in fig.(\ref{ev-r1-r2-sp-krotor-qdmodel}). The two-eigenvalue correlation function and spacing distribution exhibit the universality of (\ref{r2}) and (\ref{spacing-haake}) respectively.
 
We propose a random matrix model for the above dissipative system. We define the evolution operator, $\mathcal F$ as,
\begin{equation}
\mathcal F(\eta) = \mathcal F(0) e^{-\eta \mathcal M^2}
\end{equation}
where $\mathcal F(0)$ is a $CUE$ type matrix and $\mathcal M$ is the perturbation matrix belonging to the Gaussian unitary ensemble $(GUE)$. $\eta$ controls the dissipation. The spectral density for $\mathcal F(\eta)$ is qualitatively similar to the above mentioned dissipative kicked rotor problem. See fig.(\ref{ev-r1-r2-sp-krotor-qdmodel}). We find that the two-eigenvalue correlation function and the spacing distribution follow the above universal results.

To summarize, we have considered non-gaussian random normal matrix ensembles. We have derived analytic expressions for spectral density and $n$-eigenvalue correlation functions for arbitrary weights and shown them to be stationary and universal. We have used the MC method for generating the eigenvalues and correlation functions for such ensembles.  We have proved the single ring theorem which guarantees the existence of single ring or disk patterns of eigenvalues for any potential associated with the jpd. We illustrate our results by MC calculations for the quartic, cosine and logarithmic potentials. We study the kicked rotor and its equivalent random matrix model and find same universal properties that have been observed in the above ensembles.

RP acknowledges Ashwani Kumar Tripathi and Dushyant Kumar for useful discussion. Author also acknowledges CSIR, India for financial support.

\end{document}